\documentclass[12pt]{iopart}
\usepackage{graphicx}

\begin{document}

\title[Crossing of the Cosmological Constant...]{Crossing of the Cosmological Constant Boundary - an Equation of State Description}

\author{Hrvoje \v{S}tefan\v{c}i\'{c} \footnote{On leave of absence from the Theoretical Physics Division, Rudjer Bo\v{s}kovi\'{c} Institute, Zagreb, Croatia}}

\address{Departament d'Estructura i Constituents de la Mat\`{e}ria, Universitat de Barcelona, Av. Diagonal 647,  08028 Barcelona, Catalonia, Spain}
\ead{stefancic@ecm.ub.es}
\begin{abstract} The phenomenon of the dark energy transition between the quintessence regime ($w > -1$) and the phantom regime ($w < -1$), also known as the cosmological constant boundary crossing, is analyzed in terms of the dark energy equation of state. It is found that the dark energy equation of state in the dark energy models which exhibit the transition is {\em implicitly} defined. The generalizations of the the models explicitly constructed to exhibit the transition are studied to gain insight into the mechanism of the transition. It is found that the cancellation of the terms corresponding to the cosmological constant boundary makes the transition possible.    

\end{abstract}

\maketitle

%\section{Introduction}

Among many important cosmological problems, the phenomenon of the present, late-time accelerated expansion of the universe has come to the forefront of the observational and theoretical efforts in several last years. Apart from the exciting series of cosmological observational results confirming the accelerated character of the expansion of the universe \cite{SNIa,CMB,LSS}, we are witnessing many theoretical endeavours aimed at explaining the features of the present expansion of the universe, as well as the revival of some longstanding problems in cosmology and high energy physics, such as the cosmological constant problem \cite{cc}. 

From the theoretical viewpoint there is still no decisive insight into the nature of the accelerating mechanism. However, many promising models have been proposed to explain the acceleration in the universe's expansion. Some of the interesting approaches include the braneworld models and the modifications of gravity at cosmological scales. The most studied accelerating mechanism is the existence of a cosmic component with negative pressure, a so called {\em dark energy} component. Dark energy is a very useful concept since all our ignorance on the acceleration phenomenon is encoded into a single cosmic component. It can also be very useful as an effective description of other approaches to the explanation of the acceleration of the universe. Many models of dark energy have been constructed so far, assigning to dark energy different properties. A very general classification of these models is possible with respect to the parameter $w$ of the dark energy equation of state (EOS), $p_{d}=w \rho_{d}$, where $p_{d}$ and $\rho_{d}$ refer to the dark energy pressure and energy density, respectively \footnote{Since dark energy is the only component discussed in this paper, the subscripts $d$ will be suppressed furtheron.} . The benchmark value for the parameter of the dark energy EOS is $w=-1$ which is characteristic of the cosmological constant (CC). A problem associated to the CC value predicted by high energy physics, i.e. its discrepancy of many orders of magnitude with the value inferred from observations, is notoriously difficult. Such a situation has stimulated the development of dynamical dark energy models. Some prominent dynamical models of dark energy such as {\em quintessence} \cite{Q}, {\em k-essence} \cite{k} or {\em Chaplygin gas} \cite{Chaplygin} are characterized by $w > -1$. On the other side of the CC boundary are located models of {\em phantom energy} \cite{phantom}, with the property $ w < -1$. These models are characterized by some tension between a certain favour from the observational side and certain disfavour from the theoretical side.

Many recent analyses of observational data \cite{obser}, using ingenious parametrizations for the redshift dependence $w(z)$, show that the best fit values imply the transition of the dark energy parameter of EOS from $w > -1$ to $ w < -1$ at a small redshift. This phenomenon has been referred to in literature as {\em the crossing of the CC boundary, crossing of the phantom divide or the transition between the quintessence and phantom regimes}. It is important to stress that currently some other options, like the one of the $\Lambda$CDM cosmology, are also consistent with the observational data. Should the future observations confirm the present indications of the crossing, the aspects of the theoretical description of the crossing might provide a useful means of distinguishing and discriminating various dark energy models and other frameworks designed to explain the present cosmic acceleration. Therefore, the crossing of the CC boundary is to some extent observationally favoured and its description is a theoretical challenge.

%{\bf Work done so far}

%{\bf Crossing as a distinguishing feature?}
A number of approaches have been adopted so far to describe the phenomenon of the CC boundary crossing \cite{prethodni}. 
In our considerations of the phenomenon of the CC boundary crossing \cite{PRDcross}, we assume that that dark energy is a single, noninteracting cosmic component. We focus on the question whether the CC boundary crossing can be described using the dark energy EOS and if the answer is yes, which form the dark energy EOS needs to have to make the crossing possible. The equation of state is most frequently formulated as $p$ given as an analytic expression of the energy density $\rho$. In the considerations given below we use a much broader definition of EOS. We define the equation of state {\em parametrically}, i.e. as a pair of quantities depending on the cosmic time $(\rho(t),p(t))$, or equivalently on the scale factor $a$ in the expanding universe $(\rho(a),p(a))$. This definition easily comprises broad classes of dark energy models considered in the literature.

%\begin{equation}
%\label{eq:bianchi}
%d\rho + 3 (\rho+p) \frac{d a}{a} = 0\, 
%\end{equation}

%\begin{equation}
%\label{eq:integr}
%\int_{\rho}^{\rho_{*}} \frac{du}{f(u)} = 3 \ln \left( \frac{a_{*}}{a} \right) \, .
%\end{equation}

Let us start by considering a specific dark energy model which describes the CC boundary crossing. The dependence of the  dark energy density on the scale factor in this model is given by
\begin{equation}
\label{eq:denmod1}
\rho = C_{1} \left( \frac{a}{a_{0}} \right)^{-3(1+\gamma)} + 
C_{2} \left( \frac{a}{a_{0}} \right)^{-3(1+\eta)} \, .
\end{equation}
where $\gamma > -1$ and $\eta < -1$.
The scaling of this energy density resembles the sum of two independent cosmic components. However, we consider it to be the energy density of a {\em single} cosmic component and study its properties. Using the energy-momentum tensor conservation
%
%\begin{equation}
%\label{eq:bianchi}
%d\rho + 3 (\rho+p) \frac{d a}{a} = 0\, 
%\end{equation}
%
the expression for the dark energy pressure is obtained:
\begin{equation}
\label{eq:presmod1}
p =  \gamma C_{1} \left( \frac{a}{a_{0}} \right)^{-3(1+\gamma)} + 
 \eta C_{2} \left( \frac{a}{a_{0}} \right)^{-3(1+\eta)} \, .
\end{equation}
\begin{figure}
\centerline{\resizebox{0.55\textwidth}{!}{\includegraphics{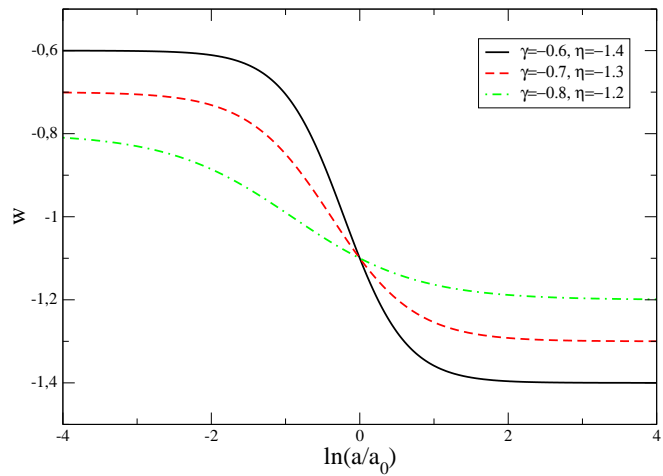}}}
\caption{\label{fig:mod1} The dark energy parameter of EOS $w$ given by (\ref{eq:wformod1}) as a function of the scale factor $a$ for the present value $w_{0}=-1.1$ and three sets of parameters $\gamma$ and $\eta$.  
}
\end{figure}
Combining (\ref{eq:denmod1}) and (\ref{eq:presmod1}) the 
expression for the parameter of the dark energy EOS acquires the following form 
\begin{equation}
\label{eq:wformod1}
w=\frac{\gamma + \eta \frac{\gamma - w_{0}}{w_{0} - \eta} \left( \frac{a}{a_{0}} 
\right)^{3(\gamma - \eta)}}{1 + \frac{\gamma - w_{0}}{w_{0} - \eta} \left( \frac{a}{a_{0}} 
\right)^{3(\gamma - \eta)}} \, .
\end{equation}
The functional dependence of the parameter $w$ on the scale factor is depicted in Fig. \ref{fig:mod1}. 
%The parameter $w$ given by (\ref{eq:wformod1}) interpolates between $\gamma$ for small values of the scale factor and $\eta$ for the  large values of the scale factor. 
The equations (\ref{eq:denmod1}) and (\ref{eq:presmod1}) can be further used to obtain 
%the direct relation between the dark energy density and pressure 
%
%\begin{equation}
%\label{eq:eosmod1}
%\left( \frac{a}{a_{0}} \right)^{-3} = \left( 
%\frac{\gamma \rho - p}{(\gamma-\eta) C_{2}} \right)^{1/(1+\eta)}
%= \left( \frac{p -\eta \rho}{(\gamma-\eta) C_{1}} \right)^{1/(1+\gamma)} \, ,
%\end{equation}
%
%which results in 
the equation of state of the studied dark energy model:
\begin{equation}
\label{eq:eosdetmod1}
\frac{p -\eta \rho}{(\gamma-\eta) C_{1}} =
\left( \frac{\gamma \rho - p}{(\gamma-\eta) C_{2}} \right)^{(1+\gamma)/(1+\eta)}
\, .
\end{equation}
The most important feature of the obtained EOS is that it is defined {\em implicitly}. This result, obtained by the explicit construction, indicates that the phenomenon of the CC boundary crossing can be studied using the implicitly defined dark energy EOS. 

Apart from the implicit character of the dark energy EOS that allows the CC boundary crossing, it would be of interest to gain additional insight into the mechanism of the crossing, i.e. the conditions necessary for the crossing to happen. To gain such an insight, we further consider the following dark energy EOS:
\begin{equation}
\label{eq:eosgenmod1}
A \rho + B p = (C \rho + D p)^{\alpha} \, ,
\end{equation}
where $A$, $B$, $C$, $D$ and $\alpha$ are real parameters. The EOS (\ref{eq:eosgenmod1}) is a generalization of (\ref{eq:eosdetmod1}) and contains it as a special case. On the other hand, another choice of parameters leads to EOS of the form $p=-\rho-K \rho^{\delta}$ \cite{PRDsing} which does not exhibit the CC boundary crossing. The generalized model (\ref{eq:eosgenmod1}) exhibits the crossing only for some parameter values and is therefore suitable for the study of the necessary conditions for the crossing. The dark energy density can be expressed in terms of parameter $w$
\begin{equation}
\label{eq:rhoexpr1}
\rho = \frac{(C+ D w)^{\alpha/(1-\alpha)}}{(A + B w)^{1/(1-\alpha)}} \, ,
\end{equation}
which leads to the equation of evolution of $w$ with the scale factor $a$:
\begin{equation}
\label{eq:evolw1}
\left( \frac{\alpha}{(F+w)(1+w)}-\frac{1}{(E+w)(1+w)} \right) dw = 3(\alpha - 1)
\frac{da}{a} \, ,
\end{equation}
where abbreviations $E=A/B$ and $F=C/D$ have been introduced.
A closer inspection of this equation reveals that there are several important values of $w$ determining its evolution: $w=-1$, $w=-F$ and $w=-E$. Whenever these values {\em exist} in the description of the problem, they represent boundaries which cannot be crossed at a finite scale factor value and can only be approached asymptotically during the evolution of the universe. This simple observation already signals that, in order to have the transition of the CC boundary, the term corresponding to the $w=-1$ boundary has to be removed from (\ref{eq:evolw1}). 

The solution of (\ref{eq:evolw1}) for the most interesting case when $E \neq -1$ and $F \neq -1$ has the form 
\begin{equation}
\label{eq:solw}
\hspace{-1.5cm} \left| \frac{w+F}{w_{0}+F} \right|^{\alpha/(1-F)} 
\left| \frac{w+E}{w_{0}+E} \right|^{-1/(1-E)}
\left| \frac{1+w}{1+w_{0}} \right|^{1/(1-E)-\alpha/(1-F)}
= \left( \frac{a}{a_{0}} \right)^{3(\alpha-1)} \, . 
\end{equation} 
This solution indicates that each of the potential boundaries can be removed by the suitable choice of the parameter $\alpha$. The boundary at $w=-F$ is removed when $\alpha=0$ and the boundary at $w=-E$ is removed when $\alpha \rightarrow \pm \infty$. The crossing of the CC boundary becomes possible with the choice $\alpha_{\mathrm cross}=(1-F)/(1-E)$, i.e. for this value of parameter $\alpha$ the CC boundary is removed. The equation (\ref{eq:solw}) then becomes  
\begin{equation}
\label{eq:solwcross}
\left| \frac{w+F}{w_{0}+F} \right| 
\left| \frac{w+E}{w_{0}+E} \right|^{-1}
= \left( \frac{a}{a_{0}} \right)^{3(E-F)} \, , 
\end{equation}
which describes the transition from the $w>-1$ regime to the
$w < -1$ regime. 

%\begin{eqnarray}
%\label{eq:solwe1}
%&&\left| \frac{w+F}{w_{0}+F} \right|^{\alpha/(1-F)} 
%\left| \frac{1+w}{1+w_{0}} \right|^{-\alpha/(1-F)} \nonumber \\
%&\times& e^{1/(1+w)-1/(1+w_{0})}
%= \left( \frac{a}{a_{0}} \right)^{3(\alpha-1)} \, .
%\end{eqnarray}

%\begin{eqnarray}
%\label{eq:solwf1}
%&& e^{-\alpha(1/(1+w)-1/(1+w_{0}))} 
%\left| \frac{w+E}{w_{0}+E} \right|^{-1/(1-E)} \nonumber \\
%&\times& \left| \frac{1+w}{1+w_{0}} \right|^{1/(1-E)}
%= \left( \frac{a}{a_{0}} \right)^{3(\alpha-1)} \, , 
%\end{eqnarray}

%\begin{equation}
%\label{eq:E1F1}
%w = -1 + \frac{1+w_{0}}{1-3 (1+w_{0}) \ln (a/a_{0})} \, .
%\end{equation}

An additional insight into the crossing mechanism can be obtained if the Eq. (\ref{eq:solw}) is studied in the rearranged form:
\begin{equation}
\label{eq:wstar}
\frac{w+\frac{\alpha E - F}{\alpha - 1}}{(F+w)(E+w)(1+w)} dw = 3 \frac{da}{a} \, .
\end{equation}
The numerator of the expression on the left-hand side can also be written as $w-w_{*}$ where $w_{*}=-(\alpha E-F)/(\alpha-1)$. For specific values of the parameter $\alpha$, the parameter $w_{*}$ can become equal to $-F$ (for $\alpha=0$), $-E$ (for $\alpha \rightarrow \pm \infty$) or $-1$ (for $\alpha_{\mathrm cross}=(1-F)/(1-E)$). Therefore, for a specific value of the parameter $\alpha$ the terms corresponding to some of the boundaries get cancelled. This cancellation is a mathematical description of the mechanism behind the CC boundary transition.

A more general model of dark energy capable of transiting between the quintessence and phantom regimes can be constructed. We consider a model with the following scaling of the dark energy density:
\begin{equation}
\label{eq:densitymod2}
\rho = \left( C_{1} \left( \frac{a}{a_{0}} \right)^{-3(1+\gamma)/b} + 
C_{2} \left( \frac{a}{a_{0}} \right)^{-3(1+\eta)/b} \right)^{b} \, . 
\end{equation}
\begin{figure}
\centerline{\resizebox{0.55\textwidth}{!}{\includegraphics{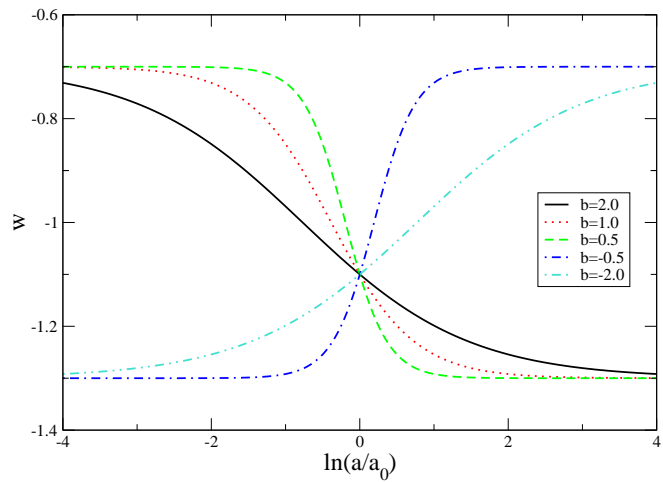}}}
\caption{\label{fig:mod2} The parameter of the dark energy EOS $w$ as a function of the scale factor of the universe $a$ for the model (\ref{eq:densitymod2}). 
The positive values of the parameter $b$ describe the transition from $\gamma$ to $\eta$ with the expansion of the universe, whereas the negative values for $b$ describe the transition from $\eta$ to $\gamma$ as the universe expands.
The values of the parameters used are $w_{0}=-1.1$, $\gamma=-0.7$ and $\eta=-1.3$.   
}
\end{figure}
This model can describe transitions of the CC boundary in both directions, i.e. from $w > -1$ to $w <-1$ and vice versa, depending on the sign of the parameter $b$, see Fig. \ref{fig:mod2}.
%For positive values of $b$ dark energy transits from the quintessence regime to the phantom regime with the expansion of the universe, whereas for the negative $b$ the transition proceeds in the opposite direction, 
 The dark energy pressure has the following form
\begin{equation}
\label{eq:pressuremod2}
p \rho^{(1-b)/b} = \gamma C_{1} \left( \frac{a}{a_{0}} \right)^{-3(1+\gamma)/b} + 
\eta C_{2} \left( \frac{a}{a_{0}} \right)^{-3(1+\eta)/b} \, .
\end{equation}
%
%and the parameter of dark energy EOS is
%
%\begin{equation}
%\label{eq:wformod2}
%w = \frac{\gamma + \eta \frac{\gamma - w_{0}}{w_{0} - \eta} \left( \frac{a}{a_{0}} 
%\right)^{3(\gamma - \eta)/b}}{1 + \frac{\gamma - w_{0}}{w_{0} - \eta} \left( \frac{a}{a_{0}} 
%\right)^{3(\gamma - \eta)/b}} \, .
%\end{equation}
%
%\begin{figure}
%\centerline{\resizebox{0.45\textwidth}{!}{\includegraphics{alphaminus15.eps}}}
%\caption{\label{fig:minus15} The dependence of the parameter $w$ on $a$ for $\alpha < (1-F)/(1-E)$. The parameter values used are $w_{0}=-1.1$, $E=0.6$ and $F=1.3$. For the used value $\alpha=-1.5$, the value of the parameter $w_{*}$ equals $-0.88$ and is situated outside the interval $(-F,-1)$.
%}
%\end{figure}
From (\ref{eq:densitymod2}) and (\ref{eq:pressuremod2}) the dark energy EOS follows directly:
\begin{equation}
\label{eq:eosdetmod2}
\frac{p -\eta \rho}{(\gamma-\eta) C_{1}} =
\rho^{((1-b)(\gamma - \eta))/(b(1+\eta))}
\left( \frac{\gamma \rho - p}{(\gamma-\eta) C_{2}} \right)^{(1+\gamma)/(1+\eta)}
\, .
\end{equation}
Starting from this explicitly constructed model, it is interesting to study its generalization in the form
\begin{equation}
\label{eq:eosmodel2}
A \rho + B p = (C \rho + D p)^{\alpha} (M \rho + N p)^{\beta} \, ,
\end{equation}
where $A$, $B$, $C$, $D$, $M$, $N$, $\alpha$ and $\beta$ are real coefficients and study the conditions for the CC boundary crossing within this generalization. Following the similar procedure as in the case of model (\ref{eq:eosgenmod1}), 
%the expression for the dark energy density becomes
%
%\begin{equation}
%\label{eq:rhomod2}
%\rho=\left( \frac{(C + D w)^{\alpha} (M + N w)^{\beta}}{A + B w} \right)^{1/(1-\alpha-\beta)} \, ,
%\end{equation}
%
%while 
the evolution law for the dark energy parameter of EOS acquires the form
\begin{equation}
\label{eq:wmod2}
\left( \frac{\alpha D}{C+D w}+\frac{\beta N}{M+N w}-
\frac{B}{A+B w} \right) \frac{dw}{1+w}
= 3(\alpha+\beta - 1) \frac{da}{a} \, .
\end{equation}
The solution of this equation in the most interesting case $A \neq B$, $C \neq D$ and $M \neq N$ is
\begin{eqnarray}
\label{eq:solwmod2}
& &\hspace{-2.0cm} \left| \frac{C+D w}{C+D w_{0}} \right|^{-\alpha D/(C-D)}
\left| \frac{M+N w}{M+N w_{0}} \right|^{-\beta N/(M-N)}
\left| \frac{1+w}{1+w_{0}} \right|^{\alpha D/(C-D)+\beta N/(M-N)-B/(A-B)} \nonumber \\
& \times & \left| \frac{A+B w}{A + B w_{0}} \right|^{B/(A-B)} 
= \left( \frac{a}{a_{0}} \right)^{3(\alpha+\beta-1)} \, . 
\end{eqnarray}
This solution for the scaling of $w$ with $a$ reveals that any of the boundaries $-A/B$, $-C/D$, $-M/N$ or $-1$ can be removed by the appropriate choice of parameters $\alpha$ and/or $\beta$. Therefore, the crossing of the CC boundary is possible in this generalized model if the exponent of $|1+w|$ vanishes. This requirement can be expressed as a condition on one of the parameters $\alpha$ or $\beta$.

Using the insight gained from the studies of the models which are explicitly constructed to exhibit the transition and their generalizations, it is possible to study a model with a nontrivial implicitly defined equation of state of the form    
\begin{equation}
\label{eq:ton}
A \rho^{2n+1} + B p^{2n+1} = (C \rho^{2n+1} + D p^{2n+1})^{\alpha} \, 
\end{equation}
and to show that the dark energy model characterized by this EOS is capable of describing the CC boundary crossing. To demonstrate the possibility of the aforementioned transition within the model (\ref{eq:ton}) we show that a suitable choice of the parameter $\alpha$ can remove the $w=-1$ boundary from the problem. The evolution equation for the parameter of EOS is
\begin{equation}
\label{eq:wforn}
\frac{w^{2n+1}+(\alpha E -F)/(\alpha-1)}{(F+w^{2n+1})(E+w^{2n+1})} 
\frac{w^{2n}}{1+w} dw = 3 \frac{da}{a} \, .
\end{equation}
where $E=A/B$ and $F=C/D$. Choosing $(\alpha E - F)/(\alpha -1) = 1$ the term corresponding to the CC boundary is removed since
\begin{equation}
\label{eq:cancel}
\frac{w^{2n+1}+1}{w+1} = \xi(w) = \sum_{l=0}^{2n} (-w)^{l} \, 
\end{equation}
and the $\xi(w)$ has no real roots. The equation (\ref{eq:wforn}) acquires the form
\begin{equation}
\label{eq:cancel2}
\frac{\xi(w)}{(F+w^{2n+1})(E+w^{2n+1})} 
w^{2n} dw = 3 \frac{da}{a} \, ,
\end{equation}
which describes smooth transitions between $w=-E^{-1/(2 n +1)}$ and $w=-F^{-1/(2 n +1)}$.

Finally, let us study the possibility of the CC boundary crossing in the dark energy model defined by the following EOS:
\begin{equation}
\label{eq:eosexp}
\rho= a e^{b p/\rho} (c - p/\rho)^{\alpha} \, .
\end{equation}
The equation for the variation of $w$ with $a$ is
\begin{equation}
\label{eq:wexp}
\left( b -\frac{\alpha}{1+c} \right) \frac{dw}{1+w} - \frac{\alpha}{1+c} \frac{dw}{c-w} = -3 \frac{da}{a} \, .
\end{equation}
The choice $b=\alpha/(1+c)$ removes the CC boundary and results in the following solution for the dark energy parameter of EOS:
\begin{equation}
\label{eq:solexp}
w=c-(c-w_{0}) \left( \frac{a}{a_{0}} \right)^{-3 (1+c)/\alpha} \, ,
\end{equation}
which describes the crossing of the CC boundary. The choice of parameters $\alpha < 0$, $w_{0} < -1$ and $-1 < c < -1/3$ yields a transition from the quintessence regime to the phantom regime at a positive redshift. 

In the generalized models of this paper used to study the CC boundary crossing, special conditions need to be met for the crossing to occur. Namely, one of the model parameters needs to acquire a value determined by the other parameter values. In a sense, if a parametric space of a model is $D$ dimensional, the set of parameter values for which the transition occurs is $D-1$ dimensional. Therefore, it has been shown that the CC boundary crossing can be described in terms of EOS, but that the model parameters need to be chosen in a special way. However, a more extensive analysis of the dark energy models with a implicitly defined EOS is needed to verify if this is a general feature of these models.   

%{\bf Additional example.}
 
%{\bf Special choice of parameters?}

%{\bf Conclusions}

In conclusion, the dark energy transition between the quintessence and the phantom regimes is studied using the dark energy EOS. It is found that in models which exhibit the CC boundary crossing, the EOS is implicitly defined. Within the generalized models the crossing is possible when there is the cancellation of the terms corresponding to the CC boundary.    
The CC boundary crossing requires a special choice of model parameters and therefore the study of its aspects might be useful in discriminating the crossing in noninteracting dark energy models from the cosmological models where the CC boundary crossing is an effective phenomenon, see e.g. \cite{SiS}.

{\bf Acknowledgements.} The author acknowledges the support of the Secretar\'{\i}a de Estado de Universidades e Investigaci\'{o}n of the Ministerio de Educaci\'{o}n y Ciencia of Spain within the program ``Ayudas para movilidad de Profesores de Universidad e Investigadores espa\~{n}oles y extranjeros". This work has been supported in part by MEC and FEDER under project 2004-04582-C02-01 and by the Dep. de Recerca de la Generalitat de Catalunya under contract CIRIT GC 2001SGR-00065. The author would like to thank the Departament E.C.M. of the Universitat de Barcelona for the hospitality.

\end{document}